\documentclass[preprint,5p,twocolumn]{elsarticle}

\usepackage{amssymb}
\usepackage{amsmath}
\usepackage{hyperref}
\hypersetup{colorlinks=true,
            linkcolor=blue,
            filecolor=magenta,      
            urlcolor=cyan,
            pdftitle={An Example},
            pdfpagemode=FullScreen,}
\usepackage{tabularx}

\usepackage{lineno}            

\journal{Biochemical Engineering Journal}

\begin{document}






\newpage
\begin{frontmatter}



\title{A novel method for quantifying enzyme immobilization in porous carriers using simple NMR relaxometry}

\author[IPI,TUD]{M. Raquel Serial}
\author[ITB,UNU-INWEH]{Luca Schmidt} 
\author[IPI]{Muhammad Adrian}
\author[ITB]{Grit Brauckmann}
\author[IPI]{Stefan Benders}
\author[ITB,UNU-INWEH]{Victoria Bueschler}
\author[ITB,UNU-INWEH]{Andreas Liese}
\author[IPI]{Alexander Penn}

\affiliation[IPI]{organization={Institute of Process Imaging},
            addressline={Hamburg University of Technology}, 
            city={Hamburg},
            country={Germany}}

\affiliation[TUD]{organization={Department of Process and Energy},
            addressline={Delft University of Technology}, 
            city={Delft},
            country={The Netherlands}}

\affiliation[ITB]{organization={Institute of Technical Biocatalysis},
            addressline={Hamburg University of Technology}, 
            city={Hamburg},
            country={Germany}}

\affiliation[UNU-INWEH]{organization={United Nations University Hub on Engineering to Face Climate Change at the Hamburg University of Technology, United Nations University Institute of Water, Environment and Health },
            , 
            city={Hamburg},
            country={Germany}}
    
\begin{abstract}
Enzyme immobilization plays a crucial role in enhancing the stability and recyclability of enzymes for industrial applications. However, traditional methods for quantifying enzyme loading within porous carriers are limited by time-consuming workflows, cumulative errors, and the inability to probe enzymes adsorbed inside the pores. In this study, we introduce Time-Domain Nuclear Magnetic Resonance (TD-NMR) relaxometry as a novel, non-invasive technique for directly quantifying enzyme adsorption within porous carriers. Focusing on epoxy methyl acrylate carriers, commonly used in biocatalysis, we correlate changes in $T_\mathrm{2}$ relaxation times with enzyme concentration, leading to the development of an NMR-based pore-filling ratio that quantifies enzyme loading. Validation experiments demonstrate that TD-NMR-derived adsorption curves align closely with traditional photometric measurements, offering a reliable and reproducible alternative for enzyme quantification. The accessibility of tabletop TD-NMR spectrometers makes this technique a practical and cost-effective tool for optimizing biocatalytic processes. Furthermore, the method holds promise for real-time monitoring of adsorption dynamics and could be adapted for a wider range of carrier materials and enzymes.

\end{abstract}


\end{frontmatter}

\section{Introduction}
The use of enzymes to facilitate complex biochemical reactions is a key area of research for economically and ecologically sustainable processes. Enzymes exhibit high stereo-, regio- and chemoselectivity, simplify downstream processing, reduce the need for harsh conditions, and reduce energy consumption compared to traditional catalysts \cite{sheldon_role_2018}. In addition, enzyme-catalyzed reactions
minimize undesirable by-products and operate under mild conditions, enhancing product quality and minimizing purification steps \cite{koeller_enzymes_2001}. Despite these advantages, widespread industrial use of enzymes is often hindered by their limited stability and lack of recyclability. To address these limitations, enzymes are frequently immobilized. 

Porous carriers have gained significant attention for enzyme immobilization due to their large surface area and tunable pore structure \cite{zhou_progress_2013}. The immobilization of enzymes within these carriers improves enzyme stability by providing a confined environment that protects enzymes against heat or exposure to organic solvents \cite{liu_facile_2017}. Furthermore, this approach allows for multipoint attachment, further enhancing enzyme stability and minimizing adverse intermolecular interactions, such as those involving proteases or gas bubbles \cite{rodrigues_stabilization_2021}. Various immobilization strategies include gel matrix encapsulation, enzyme aggregate formation by crosslinking, and direct attachment to carrier materials \cite{liese_industrial_2006}.

The immobilization of enzymes plays a crucial role in industrial biocatalysis, offering enhanced stability and reusability \cite{liese_evaluation_2013}. Studies have demonstrated their effective application on sepabeads and evaluated key factors such as the limitations of mass transport and immobilization yield to optimize industrial use \cite{hilterhaus_practical_2008}. More recent research introduced a screening platform that utilizes miniature rotating bed reactors, advancing the development of immobilized enzyme catalysis \cite{kundoch_screening_2024}.

Accurate quantification of enzyme immobilization is critical for optimizing enzyme activity and overall process efficiency. Traditional methods for assessing immobilization efficiency primarily measure the residual enzyme concentration in solution after immobilization, often using photometric tests. However, these methods involve multiple washing steps and the analysis of several wash fractions, which can introduce cumulative errors and compromise reproducibility. Moreover, they provide limited insight into the spatial distribution of enzymes within the carrier. Specifically, they cannot distinguish whether the enzymes are adsorbed on the pore surfaces or merely trapped between the carrier's particles \cite{carlsson_enzymes_2014}. 

Several physical and chemical characterization techniques are often employed to address these shortcomings. Nitrogen adsorption \cite{miyahara_adsorption_2007}, X-ray diffraction \cite{miyahara_adsorption_2007}, Fourier transform infrared spectroscopy (FTIR) \cite{yanjing_adsorption_2009} and thermogravimetric analysis (TGA) \cite{moelans_using_2005} can provide valuable information about changes in the carrier surface area, pore structure, and chemical properties before and after immobilization. These methods indirectly infer enzyme adsorption but cannot confirm the specific location of enzymes within the carrier. For direct spatial localization, fluorescence microscopy \cite{suh_analysis_2004} and transmission electron microscopy (TEM) \cite{piras_3d_2011} are often used. However, these imaging methods have limitations, including invasive sample preparation, restricted penetration depth, and potential interference with enzyme activity due to labeling dyes. While solid-state NMR spectroscopy has been used to investigate enzyme-carrier interactions at the molecular level \cite{grunberg_hydrogen_2004}, it is primarily used for structural studies and is unsuitable for routine, high-throughput quantification due to its destructive sample preparation requirements.

The need for a method that is both non-invasive and capable of directly quantifying enzyme adsorption within the pore of the carrier has driven interest in alternative techniques. Among these, Time-Domain Nuclear Magnetic Resonance (TD-NMR) has emerged as a powerful tool for probing fluid dynamics and pore structures in porous materials \cite{velasco_characterization_2023, marreiros_benchtop_2021}. Unlike traditional methods, TD-NMR can directly assess the environment within the carrier pores, offering unique insights into the interactions between enzymes and the carrier material.

In this study, we present a novel application of TD-NMR relaxometry for quantifying enzyme immobilization in porous carriers. Specifically, we focus on the covalent multi-point binding of enzymes on epoxy methyl acrylate carriers, a widely used method in biocatalysis. By analyzing changes in $T_\mathrm{2}$ relaxation times at different enzyme concentrations, we establish a quantitative relationship between NMR measurements and enzyme loading. This analysis leads to the definition of an NMR pore-filling ratio, which directly quantifies the extent of pore filling as a function of enzyme concentration. Validation results show that adsorption curves derived from the NMR pore-filling ratio closely align with those obtained through traditional photometric methods across different carrier dimensions. While TD-NMR has been explored for characterizing porous materials, this study represents, to our knowledge, its first application for directly quantifying enzyme adsorption in porous carriers. The method offers a non-invasive and reproducible alternative to conventional techniques, addressing limitations such as time-consuming workflows and lack of spatial insight into enzyme distribution. Furthermore, the affordability and accessibility of tabletop TD-NMR spectrometers positions this technique as a practical tool for accelerating biocatalytic process optimization.

This research aligns with the mission of the United Nations University Hub at TUHH by advancing sustainable biotechnological processes in accordance with the UN Sustainable Development Goals (SDGs), particularly SDG 9 (Industry, Innovation, and Infrastructure) and SDG 12 (Responsible Consumption and Production). By improving enzyme immobilization techniques through TD-NMR relaxometry, this study contributes to greener, more efficient industrial biocatalysis, helping reduce energy consumption and waste generation. The development of a reliable, non-invasive quantification method enhances the scalability of enzyme-based processes, supporting the transition toward more sustainable and resource-efficient production systems.

\section{Methods}\label{experimental_section}

\subsection{Sample preparation}
\subsubsection{Enzyme Production}
L-threonine aldolases from \textit{ E. coli} strains ME9012 and GS245 were expressed using \textit{E. coli} BL21 cultures. The expression vector pET28a wt-ltaE(N)6His was utilized to promote L-threonine aldolase production. A 20\,mL preculture was cultivated in lysogenic broth (Carl Roth, Karlsruhe, Germany) at 37\,°C for 4 hours, with kanamycin used for selection. Detailed information on the protein and gene sequences is available in the supplementary materials.
The optical density (OD) of the preculture was measured at 600\,nm, once an OD of 3 was reached, the culture was transferred to a 200\,mL main culture. After additional incubation of 2\,h, the OD ranged from 0.6 to 0.8. Enzyme expression was induced by adding isopropyl-$\beta$-D-1-thiogalactopyranoside (IPTG). Induction proceeded for 16 hours at 16\,°C.
Following induction, the cells were harvested and disrupted by sonication using a Sonopuls Generator GM 2070 (Bandelin GmbH, Berlin, Germany). Sonication was performed in 3 cycles of 3\,min each,  with settings of 9 × 10\,\% at 70\,\% power, using a 3\,mm ultrasound probe. The resulting mixture was centrifuged at 15,000\,rpm for 30\,min at 4\,°C. The supernatand obtained from this step constituted the desired cell-free extract (CFE).

\subsubsection{Purification}
The enzyme was purified using His-tag and Protino Ni-NTA agarose. A 7\,mL Protino Ni-NTA Agarose column was washed with 40\,mL of water and equilibrated with 40\,mL of binding buffer. The CFE was loaded onto the column, incubated at 4\,°C for 30\,min and then washed with 40\,mL binding buffer to remove unbound proteins. The target protein was eluted with 40\,mL elution buffer in 1\,mL fractions, followed by a buffer exchange by size exclusion chromatography to remove imidazol.

\subsubsection{Immobilization}
The immobilization process was performed as followed. The carrier was washed with 4\,$\times$\,4\, mL potassium phosphate buffer (pH 8, 1\,mol/L). Immobilization was carried out at a 1:4 w/v ratio (carrier/enzyme solution) for 47 hours at 4\,°C. After immobilization the supernatant was removed, and the carrier was washed with 2\,$\times$\,1\,mL potassium phosphate buffer (pH 8, 50\,mmol/L). Supernatants and washing fractions were collected for photometric protein concentration determinations, carriers with immobilized enzymes were utilized for relaxometry measurements. The carriers utilized in this study are presented in Table \ref{table_carrier_dimensions}. 

\begin{table*}[h!]
\centering
\small
\renewcommand{\arraystretch}{1.2}  
\setlength{\tabcolsep}{10pt}       
\begin{tabular}{l c c} \hline
\textbf{Carrier / Order Number} & \textbf{Particle Diameter [$\mu$m]} & \textbf{Pore Diameter [nm]} \\ \hline
Epoxy methacrylate / ECR8204M  & 300 -- 710  & 30 -- 60 \\ 
Epoxy methacrylate / ECR8215M & 300 -- 710  & 120 -- 180 \\\hline
\end{tabular}
\caption{\label{table_carrier_dimensions} Carriers used for immobilization.}
\end{table*}

\subsection{Bradford Assay}
For the photometric determination of protein concentration of the supernatants, the Bradford assay was applied. The measurements were performed in the microplate reader  Tecan Infinite M1000 PRO (Tecan Trading AG, Switzerland). For measurement 50\,$\mu$L of the sample is mixed with 200\,$\mu$L of bradford reagent (SERVA Electrophoresis GmbH, Heidelberg, Germany), incubated for 5\,min at room temperature, and measured at 595\,nm and 450\,nm. These values were divided (595\,nm/450\,nm), and the resulting ratio was utilized to determine the protein concentration based on a \textit{bovine serum albium} (BSA) calibration.

\subsection{NMR Relaxometry Measurements}
NMR relaxometry measurements were performed on a Magritek Spinsolve spectrometer operating at a proton frequency of 60\,MHz, at ambient temperature. Enzyme-immobilized carrier particles (approximately 50\,mg) were packed into a custom sample holder and sealed to prevent solvent evaporation during measurements. The aim was to closely replicate the conditions typically used in enzyme immobilization processes. Each sample was prepared in triplicate and measured in independent experiments to ensure reproducibility.

\begin{figure}[ht]
    \centering
    \includegraphics[width=1.0\linewidth]{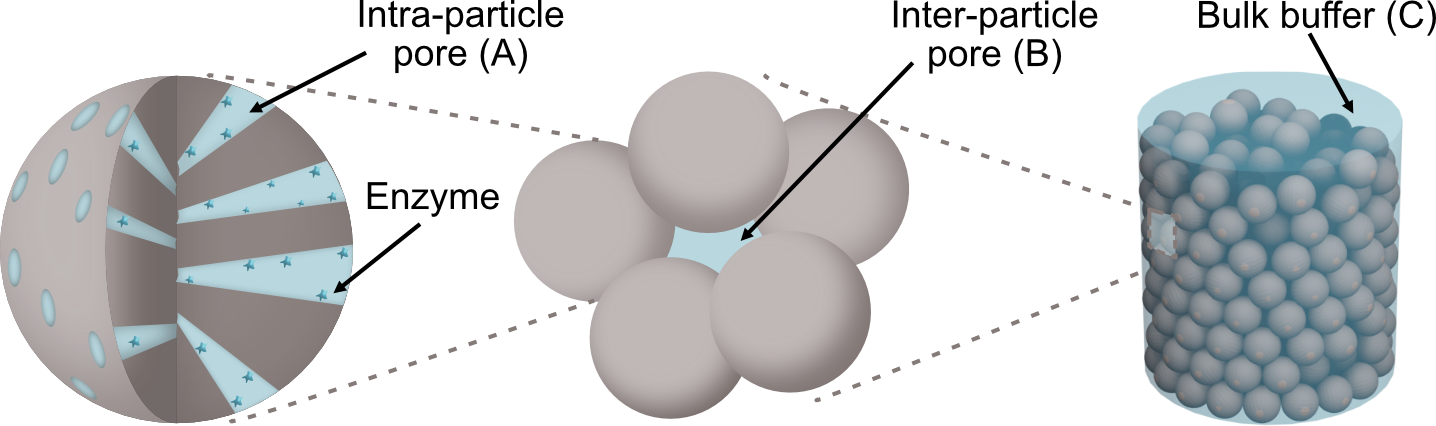}
    \caption{Schematic representation of the three solvent environments in the carrier packings used for enzyme immobilization: (A) inner-particle water confined within the pores of individual carrier particles, (B) inter-particle water located in the spaces between carrier particles, and (C) bulk water outside the carrier matrix.}
    \label{fig:scheme_carrier_packing_pores}
\end{figure}

The $T_\mathrm{2}$ relaxation times were measured using a Carr-Purcell-Meiboom-Gill (CPMG) \cite{carr_effects_1954} pulse sequence, with an echo time ($t_\mathrm{E}$) of 0.2\,ms and a total of 20000 echoes. The NMR signal decay was first analyzed qualitatively by fitting it to a continuous distribution of exponential decay functions, using an Inverse Laplace Transform (ILT) algorithm \cite{teal_adaptive_2015, serial_flintpy-nmr_2024}. Quantitative analysis was then performed using a three-exponential fitting model, which accounts for contributions from three distinct proton populations. These populations arise from different environments in the sample: (i) the intra-pore proton population (population A) associated with the solvent and adsorbed enzymes within the pores of the carrier particles, (ii) the solvent trapped between the particles (population B), and (iii) the free solvent located outside the particle packing (population C), as illustrated in Fig. \ref{fig:scheme_carrier_packing_pores}.

\section{Results and discussions}

\subsection{Relaxation dependence on enzyme concentration}
\begin{figure}[ht!]
    \centering
    \includegraphics[width=0.8\linewidth]{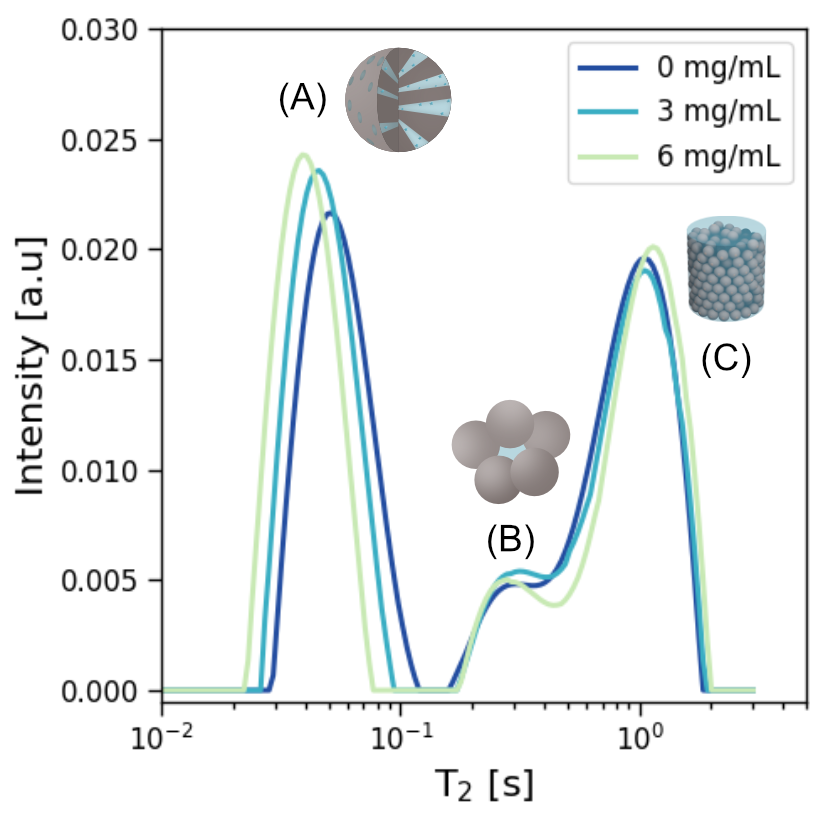}
    \caption{$T_\mathrm{2}$ relaxation time distributions for ECR8204M carrier at three different enzyme adsorption concentrations, showing populations A, B and C.}
    \label{fig:t2_ilt}
\end{figure}

Figure \ref{fig:t2_ilt} shows the $T_\mathrm{2}$ distribution for the ECR8204M carrier at three different enzyme concentrations. This distribution reveals three distinct proton populations, each corresponding to a different pore environment within the carrier's packing structure, as illustrated in Fig. \ref{fig:scheme_carrier_packing_pores}. The three populations are: (1) intra-particle protons ($T_\mathrm{2,A}$), which are located within the inner pores of the carrier particles; (2) inter-particle protons ($T_\mathrm{2,B}$), representing the buffer solution trapped between the carrier particles; and (3) bulk protons ($T_\mathrm{2,C}$), which correspond to the buffer solution outside the particle packing. The $T_\mathrm{2}$ relaxation time indicates the mobility of protons in these different environments. Protons in confined spaces, like the pores of the carrier particles (population A), experience restricted motion, leading to shorter relaxation times. In contrast, protons in bulk environments (population C), where they can move freely, exhibit longer relaxation times.

At varying enzyme concentrations, the $T_\mathrm{2}$ values for populations B and C remain largely unchanged, suggesting that these environments are unaffected by enzyme adsorption. In contrast, the $T_\mathrm{2}$ values for population A, which correspond to the intrapore protons, decrease systematically as the enzyme concentration increases. This decrease indicates that the adsorption of enzymes onto the inner pore walls restricts proton mobility, resulting in shorter relaxation times. This effect is expected, as the protons become more confined within the pores due to the presence of the adsorbed enzymes.

\begin{figure}[ht]
    \centering
    \includegraphics[width=0.9\linewidth]{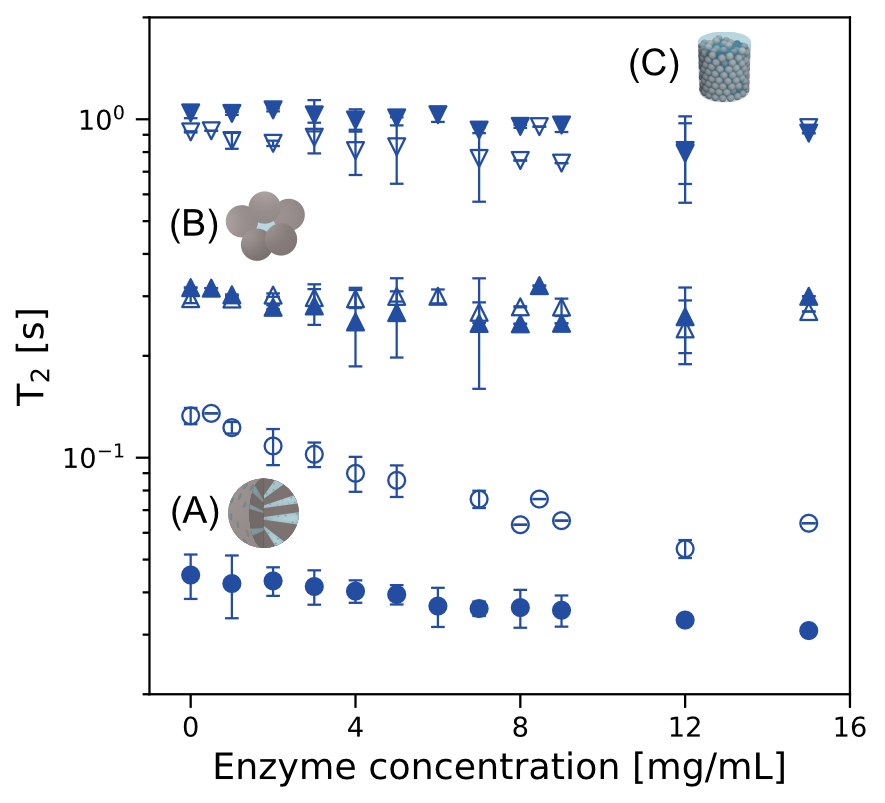}
    \caption{$T_\mathrm{2}$ relaxation time values for populations A (circles), B (upright triangles) and C (inverted triangles) as a function of solution enzyme concentrations, for carriers ECR8204M (filled symbols) and ECR8215M (empty symbols).}
    \label{fig:t2_vs_conc}
\end{figure}

The relationship between enzyme concentration and $T_\mathrm{2}$ values is further illustrated in Fig. \ref{fig:t2_vs_conc}. The $T_\mathrm{2,A}$ values for two carriers, ECR8204M and ECR8215M, show a consistent decrease with increasing enzyme concentration. The carriers used differ in their pore sizes, while their overall particle diameters remain constant (see Table \ref{table_carrier_dimensions}). This trend supports the assignment of $T_\mathrm{2,A}$ to the intra-pore protons, as the adsorption of enzymes onto the inner pore walls restricts proton mobility. In contrast, $T_\mathrm{2,B}$ and $T_\mathrm{2,C}$ remain relatively constant across the enzyme concentration range, corroborating their association with the inter-particle and bulk regions, respectively. The observed variability in $T_\mathrm{2,B}$ and $T_\mathrm{2,C}$ is likely influenced by differences in the packing arrangement and particle sizes, which affect the spatial distribution of the inter-particle and bulk regions.

\subsection{The NMR pore-filling model} 

The observed decrease in $T_\mathrm{2,A}$ in Fig. \ref{fig:t2_vs_conc} can be explained by an increase in the pore surface-to-volume ratio ($S/V$) during enzyme adsorption. For a liquid confined within a given pore of the carrier, the transverse relaxation rate $T_\mathrm{2,p}$ is dominated by surface relaxation effects and can be expressed as \cite{maillet_dynamic_2022}:
\begin{equation}
    \frac{1}{T_\mathrm{2, p}} = \rho_\mathrm{s} (\frac{S}{V})_\mathrm{pore}
\end{equation}
where $\rho_\mathrm{s}$ is the surface relaxivity, assumed constant for the carrier material. Bulk relaxation is not included in this expression, as it can be neglected for nanometer-scale pores where surface effects dominate. For cylindrical pores with a uniform radius $r_\mathrm{p}$, this simplifies to:

\begin{equation} \label{eq_T2_vs_rp}
\frac{1}{T_\mathrm{2,p}}=2\frac{\rho_\mathrm{s}}{r_\mathrm{p}} 
\end{equation}

According to this equation, $T_\mathrm{2,p}$ increases with pore radius. As enzymes adsorb onto the inner surfaces of the pores, they reduce the effective pore radius, which leads to a corresponding decrease in $T_\mathrm{2,p}$, as observed in Fig. \ref{fig:t2_vs_conc} for population A. 

\begin{figure}
    \centering
    \includegraphics[width=0.85\linewidth]{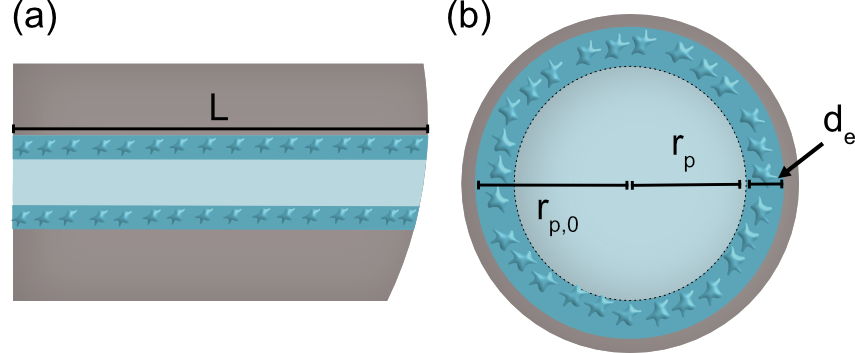}
    \caption{Schematic representation of the geometric assumptions used in the NMR pore-filling model. (a) Side view and (b) top view of a cylindrical pore within the enzyme carrier. The initial pore radius ($r_\mathrm{p,0}$) decreases to an effective radius ($r_\mathrm{p}$) after enzyme adsorption, while $L$ denotes the pore length. The model assumes a uniform enzyme layer adsorbed onto the inner pore surface, reducing the pore volume and affecting the NMR relaxation behavior.}
    \label{pore_filling_NMR_model}
\end{figure}

To quantify enzyme adsorption using NMR relaxometry, we introduce the concept of the pore-filling ratio $f_\mathrm{NMR}$, which represents the fraction of the initial pore volume that has been occupied by adsorbed enzymes:

\begin{equation}\label{eqPf_NMR}
f_\mathrm{NMR} = \frac{V_\mathrm{e,adsorbed}}{V_\mathrm{p,0}}  
\end{equation}
where $V_\mathrm{e,\text{adsorbed}}$ is the volume of adsorbed enzymes, and $V_\mathrm{p,0}$ is the initial pore volume before adsorption. Enzyme adsorption reduces the effective pore radius $r_\mathrm{p}$, which can be expressed as:
\begin{equation} 
r_\mathrm{p} = r_\mathrm{p,0}-d_\mathrm{e} 
\end{equation} 
where $r_\mathrm{p,0}$ is the initial pore radius (before enzyme adsorption) and $d_\mathrm{e}$ is the thickness of the adsorbed enzyme layer (see Fig. \ref{pore_filling_NMR_model}). For cylindrical pores of length $L$, the volume of the adsorbed enzyme is:
\begin{equation} 
\label{eqVadsorbed} 
\begin{split}
V_\mathrm{{e,\text{adsorbed}}} &= V_{p,0} - V_p \\
&= \pi L \left( r_{p,0}^2 - (r_{p,0} - d_e)^2 \right) \\
&= \pi L \left( 2r_\mathrm{p,0} d_\mathrm{e} - d_\mathrm{e}^2 \right)
\end{split} 
\end{equation}

The initial pore volume can be expressed as:
\begin{equation}
    V_\mathrm{p,0}= \pi L r_\mathrm{p,0}^2
\end{equation}

Substituting these expressions into Eq. \ref{eqPf_NMR}, the pore-filling ratio becomes:
\begin{equation} 
f_\mathrm{NMR} = \frac{(2r_\mathrm{p,0}d_\mathrm{e}-d_\mathrm{e}^2)}{r_\mathrm{p,0}^2} 
\end{equation}

To link $f_\mathrm{NMR}$ to experimentally measurable quantities, we express the enzyme layer thickness $d_\mathrm{e}$ in terms of $T_\mathrm{2, p}$ and $T_\mathrm{2,p,0}$, where $T_\mathrm{2,0}$ corresponds to the relaxation time for the unmodified pore ($r_\mathrm{p,0} = 2\rho_\mathrm{s}/T_\mathrm{2,0}$). Substituting these relationships into the above equation yields:
\begin{equation}\label{eq_final_model_expression}
f_\mathrm{NMR} = 1 - \left(\frac{T_\mathrm{2}}{T_\mathrm{2,0}}\right)^2 
\end{equation}

The final expression provides a simple, yet powerful way to estimate the extent of enzyme adsorption based solely on NMR relaxation time measurements. By measuring the $T_\mathrm{2, pore}$ at different enzyme loadings, we can derive adsorption curves that correlate enzyme concentration with pore-filling ratios.
This model offers several advantages: (1) it does not require detailed knowledge of the pore geometry or enzyme properties, (2) it eliminates the need for destructive sample processing, (3) it is adaptable to carrier particles with multimodal pore size distributions by applying the model to each pore size fraction individually. 

Finally, it is worth noting that Eq. \ref{eq_final_model_expression} is intended for detecting enzyme adsorption through NMR measurements, rather than for modeling the kinetics of enzyme adsorption. 

\subsection{Validation against photometric adsorption isotherms}

\begin{figure}[h!]
    \centering
    \includegraphics[width=0.85\linewidth]{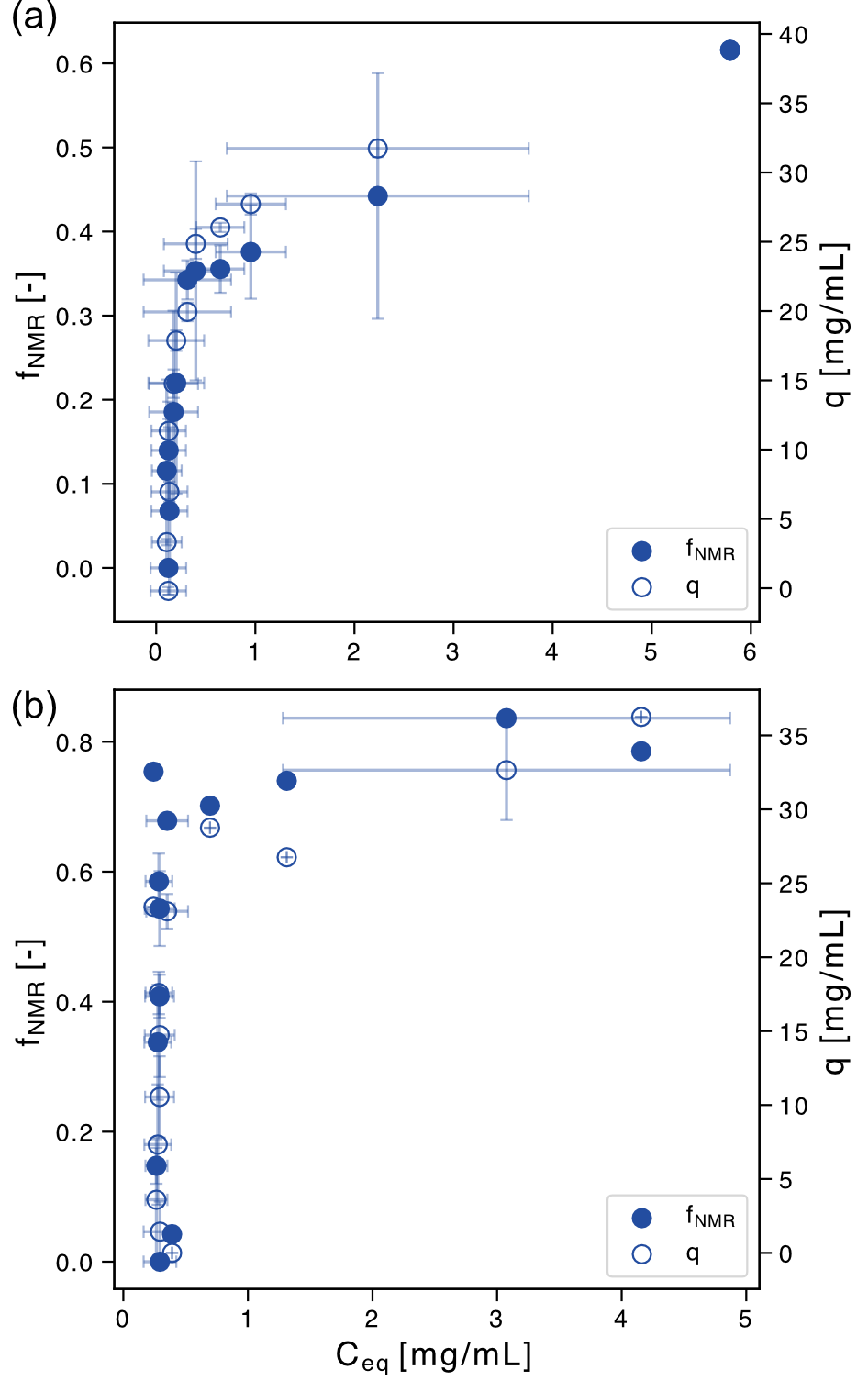}
    \caption{Comparison of NMR pore-filling fraction ($f_\mathrm{NMR}$) and photometrically determined loading ($q$) as a function of equilibrium concentration ($C_\mathrm{eq}$). Results are shown for carrier ECR8204M (a) and ECR8215M (b) after the immobilization process.}
    \label{fig:f_NMR_vs_adsorption_isotherms}
\end{figure}

Figure \ref{fig:f_NMR_vs_adsorption_isotherms} compares the calculated $f_\mathrm{NMR}$ values obtained via NMR relaxometry with the photometrically determined loading, $q$, for two different carriers ECR8204M and ECR8215M. These values correspond to different equilibrium concentrations, $c_\mathrm{equ}$, in the supernatant after the immobilization process. 

For the ECR8204M carrier (Fig.\ref{fig:f_NMR_vs_adsorption_isotherms}a), $f_\mathrm{NMR}$ values closely follow the trend of the photometric loading data, particularly at low $c_\mathrm{equ}$ concentrations. The steep initial increase in both datasets suggests that enzyme immobilization occurs predominantly via covalent attachment to the carrier surface. This behavior indicates that the ECR8204M carrier provides sufficient surface area for enzyme binding, leading to an efficient immobilization process at low concentrations. At higher $c_\mathrm{equ}$ concentrations ($\geq$ 0.5 mg/mL), both $f_\mathrm{NMR}$ and $q$ continue to increase without reaching a plateau. This trend is indicative of slower adsorption kinetics and the potential formation of adsorptive multilayers, which may result from incomplete washing steps or weakly bound enzyme layers that remain at the carrier surface \cite{carlsson_enzymes_2014, kruk_application_1997, barrett_determination_2002, vertegel_silica_2004}. These effects are consistent with the smaller pore size of ECR8204M, which limits the efficiency of the washing process.

In contrast, the ECR8215M carrier (Fig.\ref{fig:f_NMR_vs_adsorption_isotherms}b) exhibits a different adsorption behavior. Although both $f_\mathrm{NMR}$ and $q_\mathrm{max}$ a steep initial increase, a maximum loading ($q_\mathrm{max}$) is observed at approximately $c_\mathrm{equ}=$ 1 mg/mL. This difference is attributed to the significantly larger pore size of the ECR8215M carrier (approximately ten times larger than ECR8204M), which reduces the overall surface area and results in a more defined surface saturation. The washing protocol for ECR8215M appears sufficient to remove weakly adsorbed enzymes, yielding a monolayer.

The $f_\mathrm{NMR}$ data for both carriers effectively capture the trends observed in photometric adsorption measurements, including the slower adsorption kinetics and lower saturation in ECR8204M compared to ECR8215M. At higher enzyme concentrations, slight deviations are observed, but these remain within the standard deviation of the measurements.

\subsection{Stand-alone Relaxometry Method}

NMR adsorption curves offer a robust and reliable approach to monitoring enzyme adsorption. Although these curves do not directly provide enzyme loading as a function of equilibrium concentration, they closely follow the trend observed in photometric data. This makes them particularly valuable for identifying key adsorption events, such as the formation of a saturated covalent monolayer and the onset of multilayer adsorption. In contrast, conventional techniques for evaluating adsorption isotherms, which rely on equilibrium concentration measurements, are prone to inaccuracies caused by harsh immobilization conditions (e.g. enzyme precipitation at high salt concentrations). These methods also involve labor-intensive workflows and generate significant waste.

\begin{figure*}[h!]
    \centering
    \includegraphics[width=0.85\linewidth]{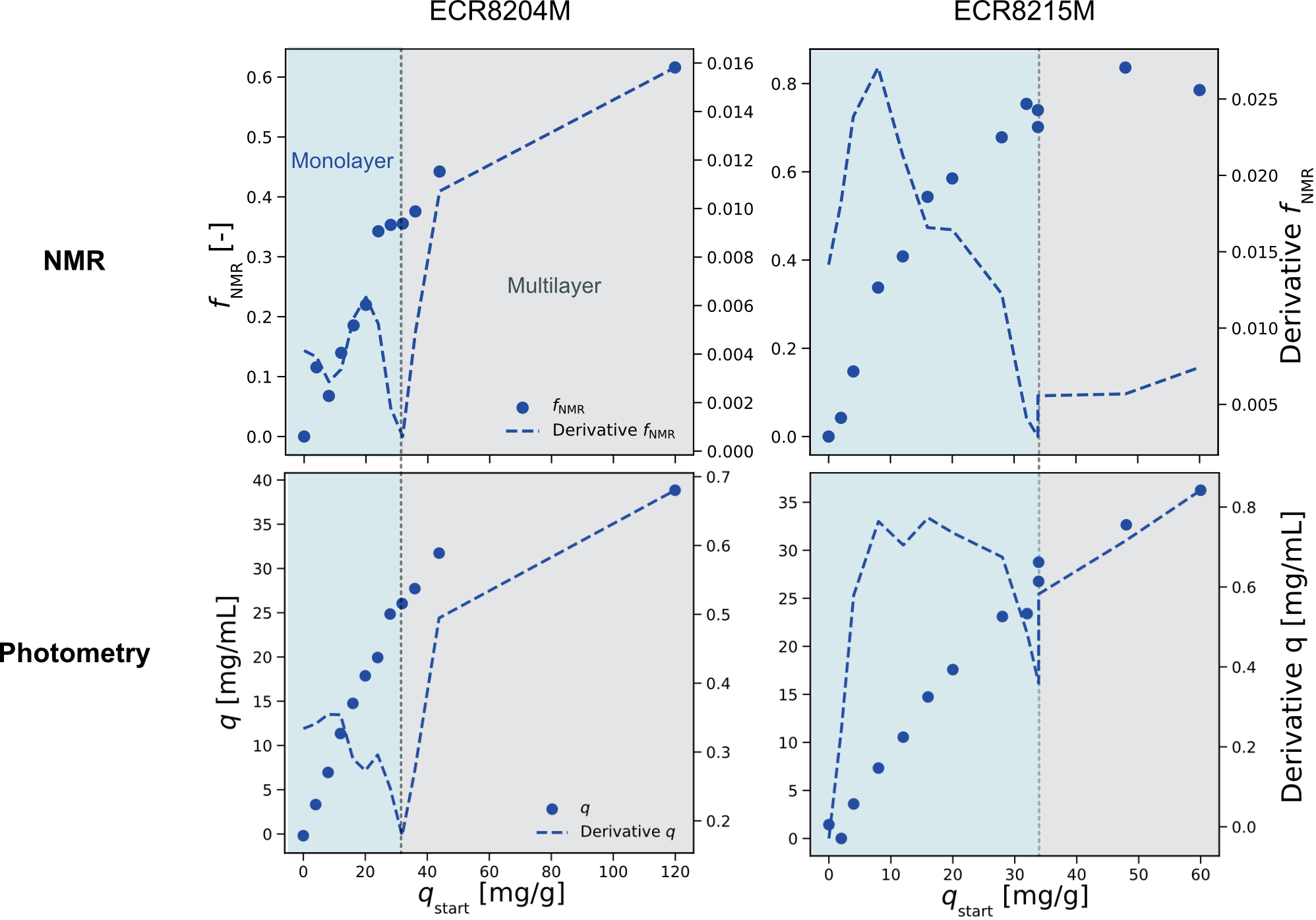}
    \caption{NMR pore-filling fraction ($f_\mathrm{NMR}$) and photometric adsorption isotherms as functions of the enzyme loading at the onset of immobilization ($q_\mathrm{start}$), along with their respective derivative plots, for ECR8204M (left column) and ECR8215M (right column). The derivative plots highlight the transitions between adsorption phases, specifically the formation of a covalent monolayer and the onset of multilayer adsorption. The inflection points in both the $f_\mathrm{NMR}$ and photometric data derivatives consistently mark these transitions for the two carriers.}
    \label{fig:standalone_method}
\end{figure*}

To address these challenges, we propose a stand-alone NMR-based method as a direct and efficient alternative for identifying adsorption transitions. The key objective of this method is to determine the fully saturated covalent monolayer during immobilization without relying on equilibrium concentration measurements. Instead, the method uses the initial enzyme concentration at the onset of immobilization ($q_\mathrm{start}$) as a reference parameter. Figure \ref{fig:standalone_method} illustrates the variation of the NMR pore-filling fraction ($f_\mathrm{NMR}$) and photometrically determined loading ($q$) as functions of $q_\mathrm{start}$ for the two carriers investigated in this study, along with their respective derivative plots. The derivative plots consistently reveal inflection points that mark the transition from covalent monolayer adsorption to multilayer formation. For carrier ECR8204M, $f_\mathrm{NMR}$ derivative plots show that the enzyme enzyme-to-carrier mass ratio required to achieve a single covalent monolayer is 31.9\,[mg/g], while for ECR8215M, the required mass ratio is 33.9\},[mg/g]. Identical results are found for photometry data. 

Despite differences in pore size and surface area, both carriers show similar enzyme concentrations per carrier to achieve a fully saturated monolayer. Interestingly, the smaller pore size of the carrier ECR8204M might have been expected to allow higher enzyme concentrations to form the monolayer. However, two factors may explain why this is not observed: (i) enzyme diffusion into smaller pores during immobilization might be restricted, preventing complete surface coverage of the carrier \cite{bayne_effect_2013}, (ii) the hydrodynamic radius and/or folding of the enzymes under specific conditions could inhibit a high number of enzymes from fitting into smaller pores, leading to a less effective immobilization process \cite{gao_effects_2012}. Further experiments are required to confirm these hypotheses and explore the interplay between pore size, enzyme conformation, and adsorption efficiency.

\subsubsection{Method limitations and perspectives}

While the proposed method effectively describes enzyme adsorption in porous carriers, several limitations should be acknowledged to guide its application and interpretation in other systems. 

One key prerequisite is the ability to distinguish protons within pores from those in the surrounding bulk solvent, as demonstrated for populations A, B and C in Fig.\ref{fig:t2_ilt} for the investigated carriers. Variations in carrier materials or enzyme properties can alter surface relaxivity $\rho_\mathrm{s}$, leading to shifts in $T_\mathrm{2}$ relaxation times. These shifts may either complicate differentiation between pore and bulk protons, for example, in cases where pore sizes are large enough to approach inter-particle pore dimensions, or result in relaxation times that are too short to measure reliably. This challenge is particularly evident in systems containing paramagnetic centers, such as metal-organic frameworks (MOFs), where very short signal lifetimes prevent the use of conventional benchtop NMR equipment \cite{velasco_modulation_2019}. Additionally, environmental factors such as temperature, pH, and buffer composition may influence surface relaxivity or enzyme interactions, introducing variability that needs to be carefully controlled or accounted for during experiments. 

The method also assumes that $\rho_\mathrm{s}$ remains consistent and comparable between the carrier and enzyme layers. However, interactions between the solvent and the carrier surface may vary significantly depending on material properties such as hydrophobicity, hydrophilicity, or coating mechanisms. Future work will focus on exploring different approaches to differentiate such components, including 2D NMR experiments such as relaxation correlation maps ($T_\mathrm{1}$-$T_\mathrm{2}$ or $T_\mathrm{2}$-$T_\mathrm{2}$) \cite{silletta_enhanced_2016, dong_toward_2024, robinson_low-field_2021} and Diffusion-$T_\mathrm{2}$ (D-$T_\mathrm{2}$) measurements \cite{terenzi_spatially-resolved_2019, zhang_spatially_2014}, which have proven to be useful in detecting bound water in porous media. 

Another consideration is the assumption of homogeneous enzyme coverage within the pores. The model presumes uniform deposition both radially and longitudinally along the pores, which may not hold true at high enzyme concentrations. Additional modeling is required to address deviations, such as uneven distributions or inconsistent binding mechanisms across pores. 

Finally, the method requires pores of sizes in which monolayer adsorption results in a measurable change in the pore surface-to-volume ratio. In systems where this condition is not met, the method's sensitivity to small changes may be limited. 

\section{Conclusions}
This study introduces a novel application of TD-NMR for quantifying enzyme immobilization within porous carriers. By monitoring changes in $T_\mathrm{2}$ relaxation times with increasing enzyme concentration, the method enables the calculation of a pore-filling fraction $f_\mathrm{NMR}$ that provides a direct measurement of enzyme adsorption. This approach addresses several limitations of traditional photometric methods.

Validation of the TD-NMR method against photometric adsorption isotherms confirms its reliability and reproducibility, particularly in the capture of critical adsorption transitions. This capability is further enhanced by the proposed stand-alone method, which, by taking the derivative of NMR adsorption curves, facilitates the precise identification of inflection points critical for optimizing immobilization protocols.

The proposed method offers practical advantages, such as minimal sample preparation, reduced material consumption, and direct and fast measurements. These features make it particularly suitable for high-throughput screening and real-time monitoring of enzyme immobilization. However, the performance of the current method is bound to several factors that may influence its applicability to other systems. For instance, variations in carrier properties, such as surface coating and enzyme interactions with the carrier surface, require further investigation and are the focus of future work.

\section{Declaration of Competing Interest}
The authors declare that they have no competing interests.

\section{Acknowledgements}

L.S. and A.L. are grateful to the "Deutsche Forschungsgemeinschaft (DFG)" for financial support (grant number: 454753456) and to Prof. Dr. Harald Gröger and Logia Jolly for supplying the plasmid for threonine aldolases synthesis. A.L And A.P. acknowledge funding by the Deutsche Forschungsgemeinschaft (DFG, German Research Foundation) – SFB 1615 – 50385073.

\section{Author contributions}
Conceptualization: M.R.S., L.S., S.B. Experimental measurements: M.A., G.B. Formal analysis and Data curation: M.R.S., L.S. Writing—original draft: M.R.S., L.S. Writing—review and editing: M.R.S., L.S., S.B., V.B., A.L., A.P. Funding acquisition: A.P., A.L. All authors have read and agreed to the published version of the manuscript.

\section{Data availability}
All data and Python scripts, including Jupyter notebooks used to process and analyze the data, are publicly available at https://collaborating.tuhh.de/v-10/public/manuscripts/.
\section{Declaration of generative AI and AI-assisted technologies in the writing process}
This paper was copy-edited with the assistance of AI technologies as ChatGPT and Grammarly AI in order to improve clarity and readability. The authors have thoroughly checked all the proposed edits and take full responsibility for the content of this manuscript.

\bibliographystyle{elsarticle-num}
\bibliography{enzyme_paper_zotero}

\end{document}